\documentclass[preprint,12pt]{elsarticle}
\usepackage{graphicx}
\usepackage{dcolumn}
\usepackage{epsfig} 
\usepackage{bm}
\usepackage{color}
\usepackage{epstopdf}

\usepackage{amssymb}
\usepackage{lineno}

\def\be{\begin{equation}}
\def\ee{\end{equation}}
\def\bea{\begin{eqnarray}}
\def\eea{\end{eqnarray}}
\def\pt{\partial}
\def\ffi{\varphi}

\def\al{\alpha}

\def\eps{\varepsilon}
\def\Dt{\Delta}
\def\bt{\beta}

\def\sign{\mbox{sign}}

\def\pt{\partial}

\newcommand{\grl}{  {Geophys. Res. Lett. }}

\newcommand{\pre}{  {Phys. Rev. E }}

\begin{document}

\begin{frontmatter}

\title{Nonlinear resonances generate large-scale vortices in the phase space in plasma systems}

\author{Fan Wu$^{1}$, Dmitri Vainchtein$^{2,3}$, Anton Artemyev$^{4,3}$}

\address{
$^{1}$ School of Traffic and Transportation Engineering, Central South University, Changsha, China\\
$^{2}$ Nyheim Plasma Institute, Drexel University, Camden, NJ, USA\\
$^{3}$ Space Research Institute, Moscow, Russia \\
$^{4}$ Institute of Geophysics and Planetary Physics, University of California, Los Angeles, CA, USA}

\begin{abstract}
It is well-known that the resonance phenomena can destroy the adiabatic invariance and cause chaos and mixing. In the present paper we show that the nonlinear wave-particle resonant interaction may cause the emergence of large-scale coherent structures in the phase space. The combined action of the drift due to nonlinear scattering on resonance and trapping (capture) into resonance create a vortex-like structure,  where the areas of particle acceleration and deceleration are macroscopically separated. At the same time, nonlinear scattering also creates a diffusion that causes mixing and uniformization in around the vortex. 
\end{abstract}


\end{frontmatter}

\section{Introduction}
Convection cells and large-scale coherent structures are common place in the physical space of fluid mechanical systems. However, the same structures in the phase space of Hamiltonian systems describing plasma settings are much more rare. In the present letter we introduce a simple system of a charged particle moving in a non-uniform magnetic field in the presence of an electrostatic wave.

In many plasma systems, charged particle resonant interaction with electromagnetic waves represent the only way of an efficient energy exchange between particle populations. For coherent resonant interaction, when resonances are not destroyed and particle motion is not chaotic, particles can spend a long time within the resonance and their dynamics become much more complicated than a diffusion in velocity space. The analysis of this dynamics is based on modelling charged particle motion in an effective potential generated by a combination of Lorentz forces from background magnetic field and wave electromagnetic field. Particles trapped into such potentials, may form large-scale vortices in the phase space. The internal structure of these vortices and their evolution due to particles exchange between trapped and nontrapped (transient) populations determine the wave dumping/growth, particle acceleration/deceleration, and many other important wave characteristics. In a classical problem of the nonlinear Landau damping, the trapped phase space region is assumed to be uniformly filled, and thus the events of particle trapping/escape can influence the wave dynamics \citep[e.g.,][]{ONeil65,Mazitov65}. In many realistic systems effects related to formation and evolution of vortices control the primary wave dynamics and secondary wave formation \citep[e.g.,][]{Veltri05, Dodin&Fisch12:III, Benisti17:I, Tao17:generation}. Vortex boundary is defined by the separatrix demarcating trapped and transient populations. Vortex evolution is determined by two competing processes: phase space mixing due to the  trapping/escape accompanied with by particle phase scattering due to separatrix crossing and regular (adiabatic) changing of trapped volume due to wave and background electromagnetic field change.

In the present paper we present a simple setting where the nonlinear resonance phenomena create a phase-space vortex. The structure of the paper is as follows. We start with the main equations and introduce the separation of time scales. Then we introduce the main resonance and describe scattering at resonance and trapping (capture) into resonance at a single crossing. The we estimate the time-scale required for total mixing of particles and for a global turn of the vortex. Finally, we use propose a kinetic equation that describes the leaking of particles from the vortex.

\section{Main equations}

Consider the following system that corresponds to a charged particle moving in a double-well potential in the presence of a fast wave.

A dimensionless Hamiltonian of a particle can be written as
\bea
H &=& \frac12 p_y^2 + U(y) + \bt\sin\chi\ffi
\label{1} \\
U(y) &=& \left( \al^2 y^2 -1\right)^2/8\al^2, \quad \ffi = y-ut \nonumber
\eea
Here $\chi$, $u$, and $\bt$ are the wavelength, the phase speed, and the amplitude of the wave, respectively. In what follows we assume that the wave is short, $\chi = 30 \gg 1$, and weak, $\bt = 0.1 \ll 1$. The other parameters are $\al=0.1$ and $u=1$.

\begin{figure}[ht]
\centering
\includegraphics[width=2.3in]{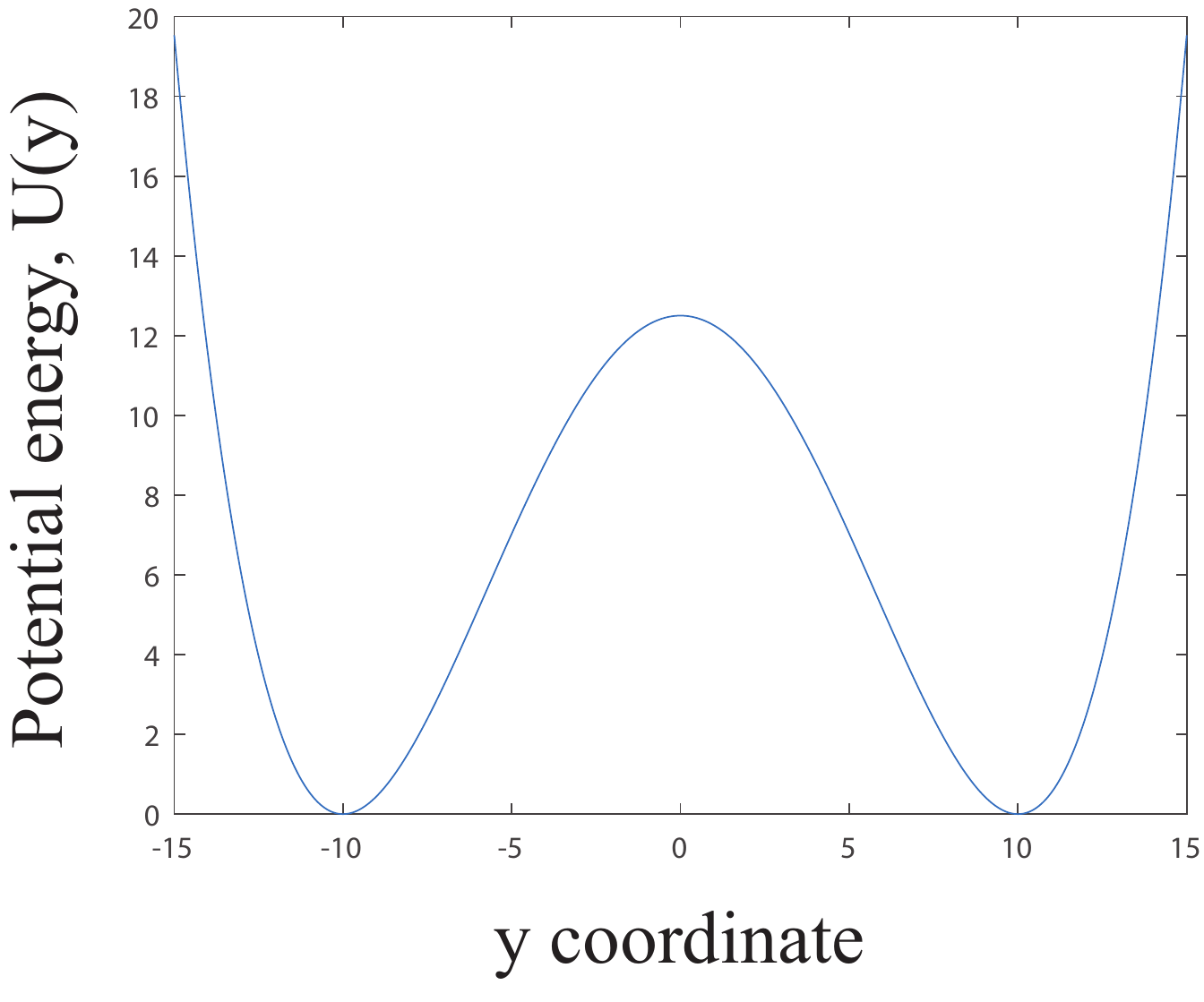}
\caption{\label{f1} Profile of the potential energy for the unperturbed system.}
\end{figure}
On the energy plot, $H_{min}=0$ is the minimum possible value of energy (at the bottom of the either of the two potential wells), $H_C=1/(8\al^2) = 12.5$ is the value of energy at the potential barrier at $y=0$.

\section{Separation of time scales and method of averaging. Adiabatic invariants}

The smallness of the parameter $1/\chi \ll 1$ introduces a separation of time scales. The variable $\ffi$ is fast, and the variables $(x,p_x)$ and the energy $H$ are slow. In the first approximation we can average Hamiltonian (\ref{1}) over the fast phase, which is effectively equivalent to omitting the term $\bt\sin(\chi \ffi)$ in (\ref{1}):
\be
H_{av} = p_y^2/2 + \left( \al^2 y^2 -1\right)^2/8\al^2
\label{2a}
\ee
The averaged Hamiltonian $H_{av}$ does not depend on time explicitly, therefore $H_{av}$ is an integral of the averaged system. In the exact system (\ref{1}) the value of $H$ is approximately conserved (with the accuracy of order $\bt$) everywhere, where the separation of time scales is valid and the method of averaging works. The averaging fails where the the rate of change of $\ffi$ vanishes on the line called a {\it Resonance}:
\be
d \ffi/d t = p_y - u = 0
\label{3}
\ee

The particle's dynamics is drastically different depending on it the particle intersects the resonance or not. The location of the resonance is defined wave phase velocity $u$. When $u$ is large, particles in the bottom wells do not come to the resonance, and their dynamics remains regular even in the presence of the wave.

When $u$ is smaller, on each period of the fast motion, most of the particles cross the resonance twice. There are two kind of exceptions: the trajectories near the bottom of the two wells that do not cross the resonance at all, and the trajectories just above the separatrix that cross the resonance four times. Every time a particle crosses a resonance, the value of energy $H$ changes. There are two main phenomena occurring at the resonance: capture (trapping) into resonance and scattering on resonance, see \citep[e.g.,][]{Artemyev18:cnsns} and references therein.

\section{Jump of energy at the scattering}

During most of the resonance crossings, the energy of a particle changes only slightly. This process is called scattering on resonance. It follows from \citep[e.g.,][]{Neishtadt11:mmj, Artemyev18:cnsns} and references therein that the change of the energy is given by
\be
\Delta H = -u\sqrt{\bt}\sqrt{2a}\int\limits_{-\infty}^{\tilde \varphi_*} \frac{\cos \tilde \varphi \; d\tilde \varphi}{\sqrt {2\pi\xi + \tilde \varphi - a \sin \tilde \varphi} }
\ee
where $\tilde \varphi = \chi \varphi$, $a= \beta \chi/A$ and
\be
A = A(H) = \frac{\pt U}{\pt y} = \frac12 y_R \left( \al^2 y_R^2 - 1 \right)
\ee
The value of $A$ is computed at the resonance crossing $y = y_R (H)$, given by the condition
\be
H = \frac12 u^2 + \left( \al^2 y_R^2 -1\right)^2/8\al^2
\ee
Here $\tilde \varphi_*$ is the value of $\tilde \varphi$ at the moment of crossing of the resonance, $2\pi\xi = \tilde \varphi_* - a \sin \tilde \varphi_*$. The value of $\xi$ is a very sensitive function of the initial conditions: even small, order $\eps$, changes in the initial conditions result in significant changes in $\xi$, see \citep[e.g.,][]{Artemyev18:cnsns} and references therein. Therefore, for multiple consecutive scattering $\xi$ can be treated as a random variable uniformly distributed on $(0,1)$, see a numerical verification of this assumption in \citep{Itin00}). Correspondingly, $\Delta H$ becomes a random variable as well.

Statistical properties of $\Delta H$ depend on the value of $a$. When $a>1$, the average value of $\Delta H$, defined as
\be
\left< \Delta H \right> = \int_0^1 \Delta H(\xi) \; d\xi
\ee
is finite \citep{Neishtadt99}. Alternatively, when $a<1$, $\left< \Delta H \right> =0$. The second moment of $\Delta H$ is always finite:
\be
\left< \left( \Delta H \right)^2 \right> = \int\limits_{0}^{2\pi} \left(\Delta H(\tilde \xi) - \left< \Delta H \right>\right)^2 d\xi
\ee

\section{Capture (trapping) into resonance}

Besides scattering, particles may be trapped (captured) into resonance. Trapped particles move for a while with the wave. Trapping is possible if $a>1$ and along the trajectory $da/dt>0$. It follows from equations from the Hamiltonian and the definition of $a$, that particles can be captured on the left walls in both of the wells and released from capture on the right walls.

It was shown in \citep[e.g.,][]{Neishtadt75,bookAKN06,Artemyev18:cnsns} that trapping can be considered to be a random process. While for every given particle approaching the resonance it can be predicted will it be trapped or not (provided trapping is possible), this trapped-or-not-trapped transition is very sensitive to initial conditions. Therefore, for multiple particles and multiple passages through resonance it is reasonable to consider trapping as a probabilistic process. The method of computing the probability of trapping, $\Pi(a)$ is presented in several papers, see \citep[e.g.,][]{Artemyev15:pop:probability, Artemyev18:cnsns} and references therein.

\begin{figure}[ht]
\center\includegraphics[width=2.3in]{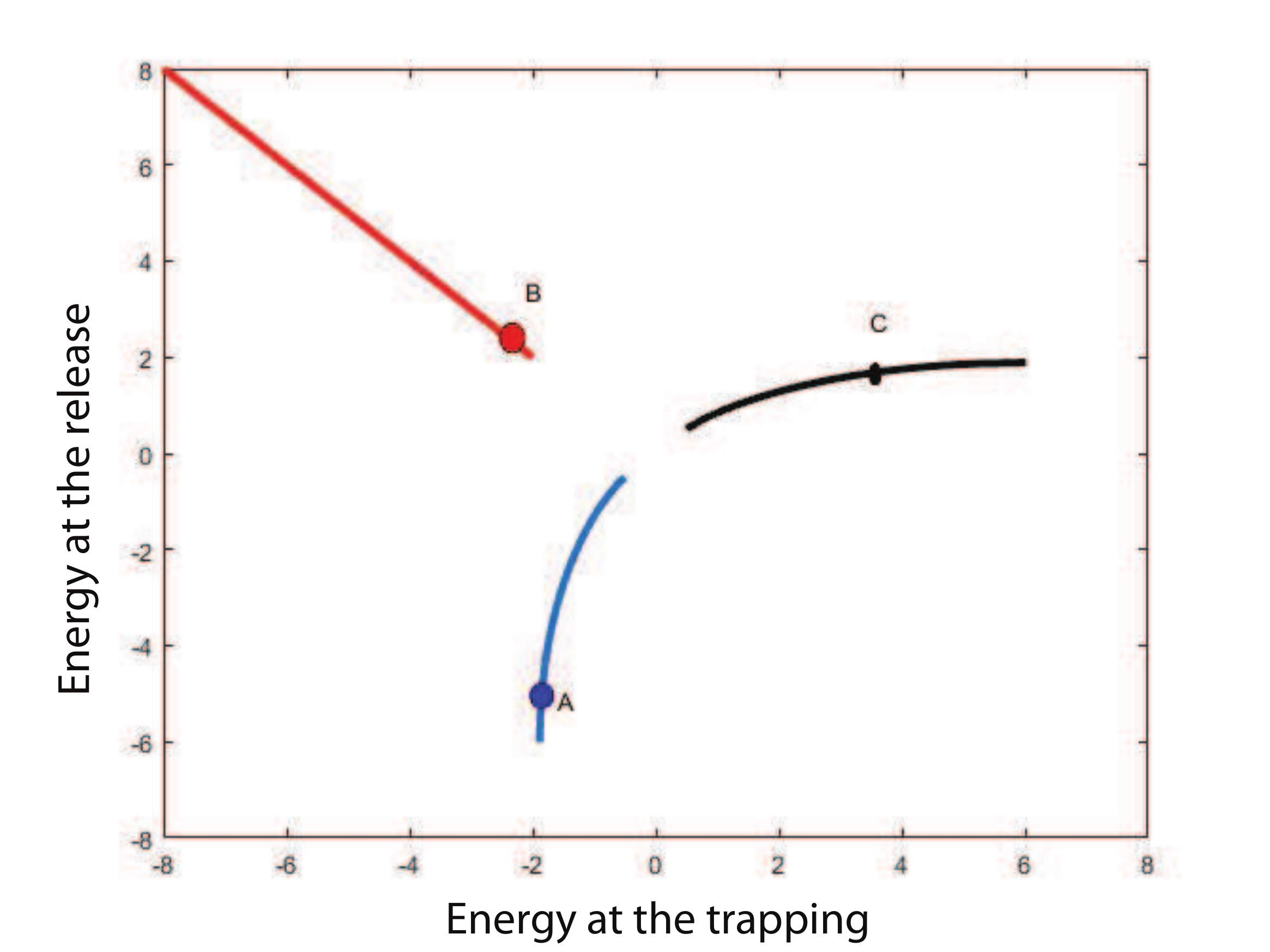}
\caption{\label{f6} Trapping (capture) into resonance: input-output function. Horizontal axis: $H\sign(y)$ at the trapping, vertical axis: $H\sign(y)$ at the release.}
\end{figure}

Once a particle is trapped into resonance, its dynamics is integrable and predictable. In particular, the value of the energy a the release from resonance can be computed explicitly, \citep[e.g.,][]{Artemyev17:pre}. The energy input-output function is presented in Fig.~\ref{f6}. When plotting Fig.~\ref{f6}, we used the quantity $H\sign(y)$ instead of $H$ to distinguish between the left and the right wells.

In the left well, the location where the trapped particles are transported to depends on the energy at which the trapping occurred. For $H_{tr} > H_{lr}=2$, particles are transported to the right wall of the right well (L-R capture). From the symmetry of the potential well with respect to the axis $x=0$, in this case the energy does not change. This trapping corresponds to the red line in Fig.~\ref{f6}. For $0<H_{tr} < H_{lr}$, particles are transported to the right wall of the left well (L-L capture). In this case, the energy grows (the blue line in Fig.~\ref{f6}).

In the right well, all the particles captured at the left wall are transported to the right wall. The energy decays (the black line in Fig.~\ref{f6}).

\begin{figure}[ht]
\centering
\includegraphics[width=5in]{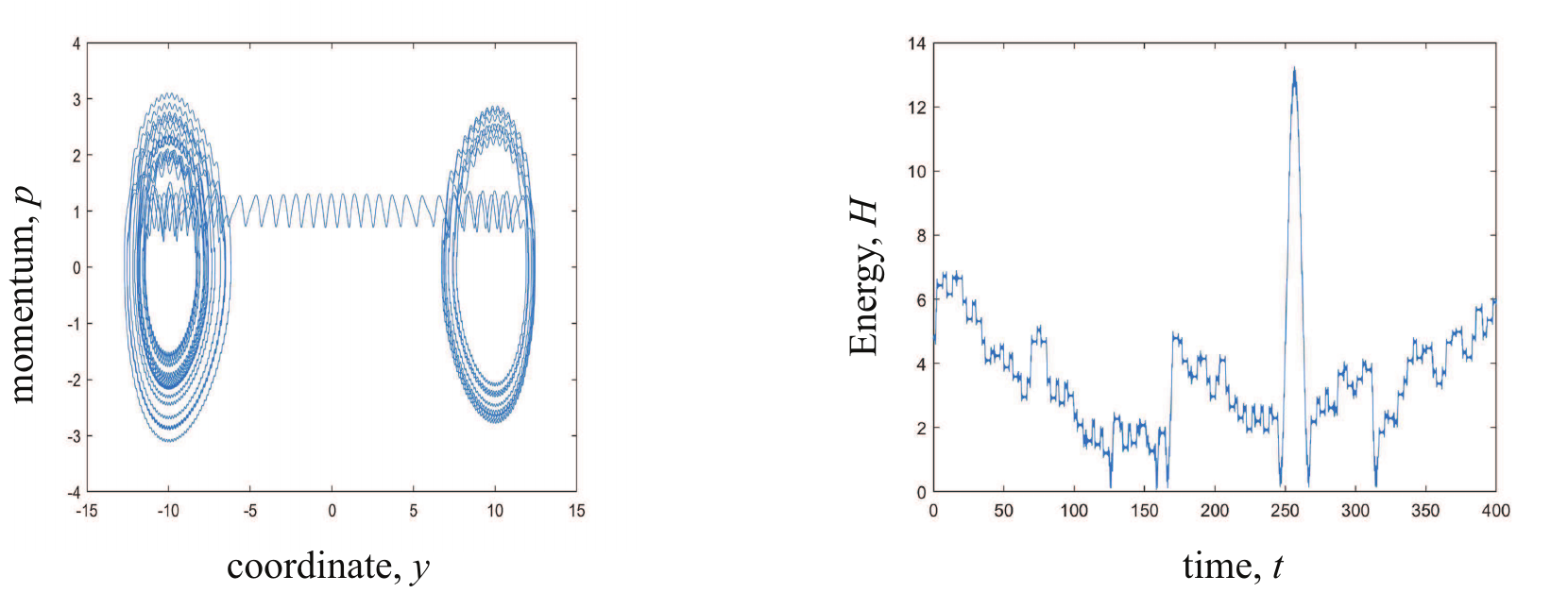}
\caption{\label{f5} Trapping (capture) into resonance. (a) Phase portrait; (b) Energy evolution.}
\end{figure}

An interval of the phase trajectory containing all three types of trapping is presented in Fig.~\ref{f5}. First the particle is trapped (at about $t=170$ in the right panel in Fig.~\ref{f5}, the outer spiral in the left well in the left panel in Fig.~\ref{f5}) in the left well and released in the left well (Point A in Fig.~\ref{f6}). The second time the trapping occurs at a higher value of the energy (at $t\approx 245$ in the right panel in Fig.~\ref{f5}, the long spiral in the left panel in Fig.~\ref{f5}), and the particle is transported to the right well (Point B in Fig.~\ref{f6}). The third trapping (at $t\approx 310$ in the right panel in Fig.~\ref{f5}, the outer spiral in the right well in the left panel in Fig.~\ref{f5}) move the particles from the left wall of the right well to the right wall (Point C in Fig.~\ref{f6}).

\section{Medium-time description: Phase space vortex}

\begin{figure}
\centering
\includegraphics[width=5in]{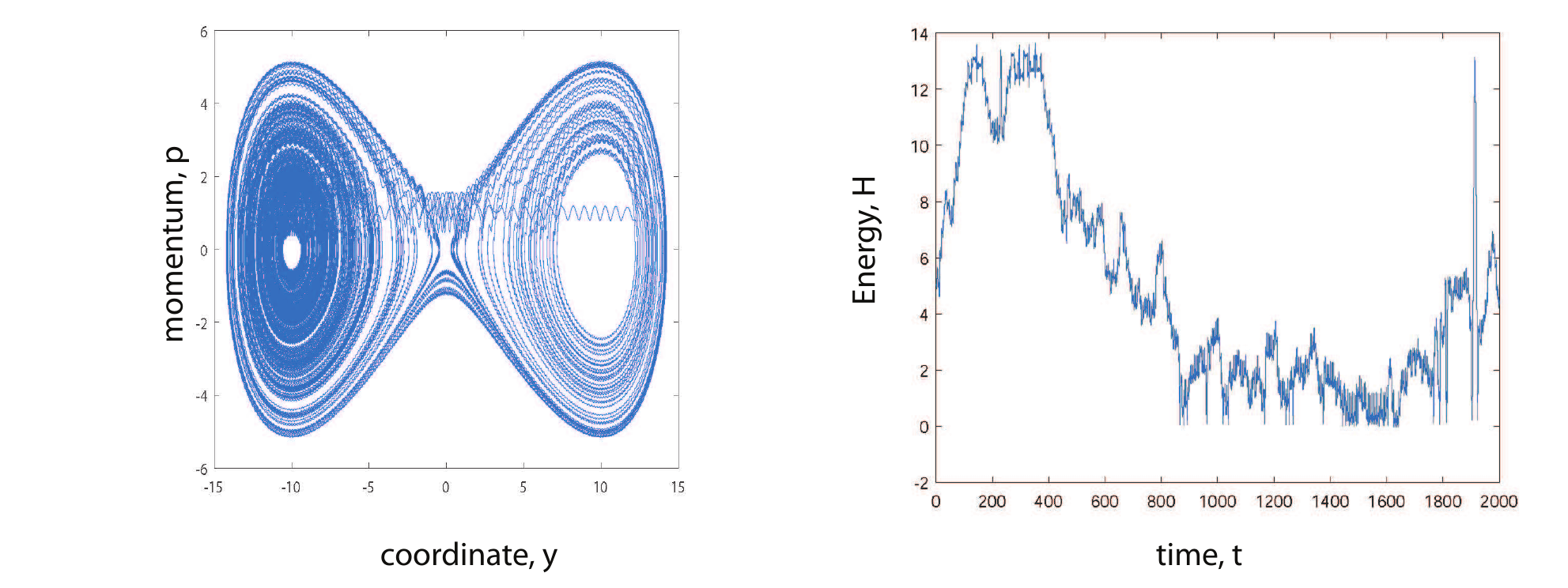}
 \caption{\label{f2} Medium-time evolution of a single particle from the bottom of the right well, up to the top well, down the left well, and transfer by capture into resonance to the right well, and starting up again. Left panel: phase plane. Right panel: Energy evolution.}
\end{figure}

The medium-time behaviour of any given particle consists of successive motion in three distinct domains: the right well, the left well, and the top (above the barrier). This dynamics is illustrated in Fig.~\ref{f2}. In the right wall, the value of energy grows on average, thus we have a drift towards large values of energy. Then particles enter the top domain (at about $t\approx 200$) where there is no drift, and we have pure diffusion. In terms of energy, particles can go up -- where there is no bound -- or down. Most of the particles stay some time at the top domain ($t\approx 200-400$). If they go down to the right well, advection kick them to the top domain immediately. However, if a particle goes down the left domain ($t\approx 400$), advection takes it, and drags down the left well ($t\approx 400-900$). Towards the bottom of the left well, trapping into resonance becomes possible. While the energy is not too small, above $H_{lr}$, trapping into resonance takes the particles from the left well to the right well: the L-R trapping. Once below the $H_{lr}$, trapping keeps the particles inside the left well: the L-L trapping. However, what is important, is that the increase of energy in the L-L trapping the energy at the release from resonance is larger then $H_{lr}$. Therefore, the L-R trapping becomes possible again. Particles oscillate near the bottom of the left well undergoing downward drift and upward L-L trappings ($t\approx 900-1900$) until finally the the L-R trapping occurs (($t\approx 1900$)). After that particle is transported to the right well and the process repeats again.

\begin{figure}
 \centering
\includegraphics[width=2.3in]{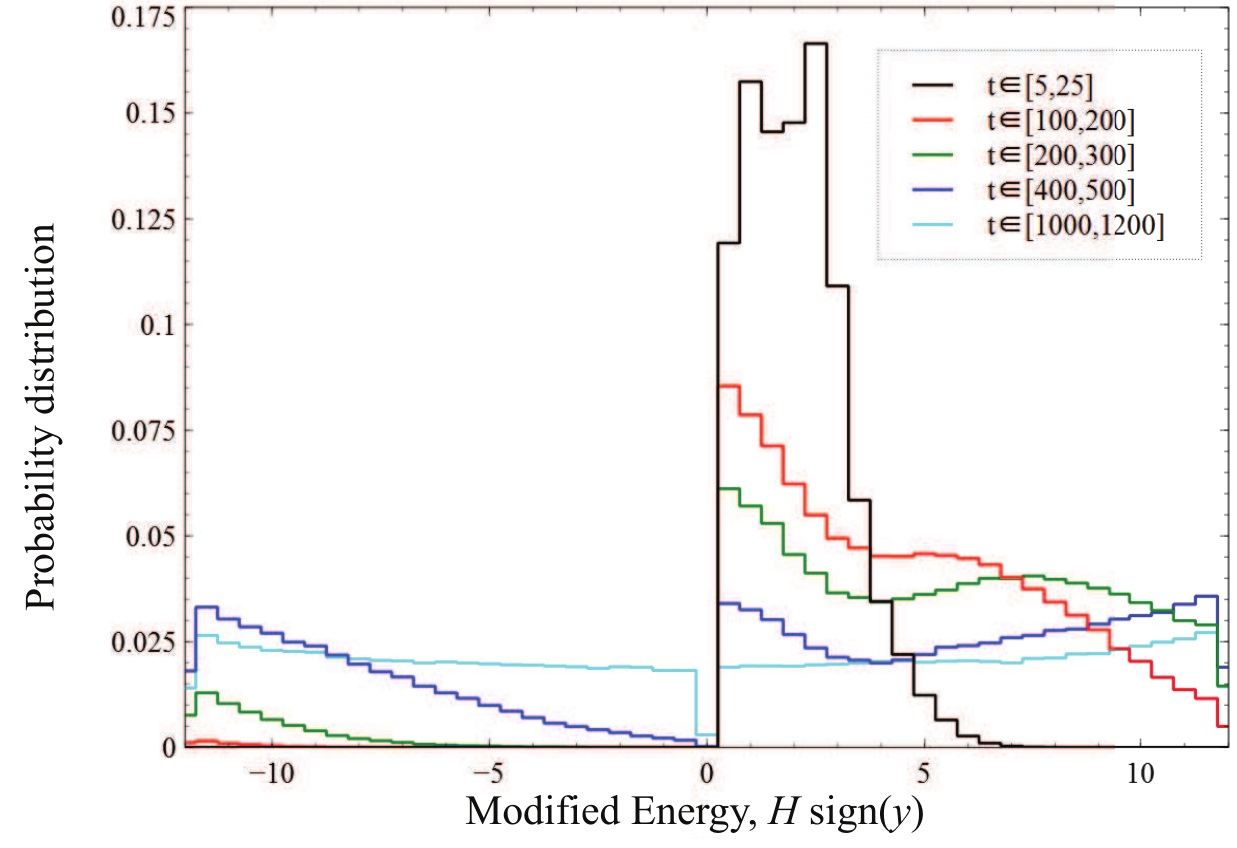}
 \caption{\label{f7} Medium-time evolution of a particle ensemble: histograms of the PDF for different values of time. From the initial distribution in the right well (black curve), over the top, to the left well. Note, that the horizontal axis is the energy H multiplied by $\sign(y)$. So the left and the right wells corresponds to $H\sign(y) <0$, and $H\sign(y) >0$, respectively. }
\end{figure}

To describe the evolution of the ensemble of particles, we introduce a Probability Distribution Function (PDF). We performed a set of numerical simulations of $20,000$ particles that were originally localized at $H=1$ in the right well, Fig.~\ref{f7}. Particles start as a relatively narrow distribution in the right well (black curve). Scattering on resonances cause the maximum of the curve to move to the right, while the distributions become wider. The particle start arriving to the top of the hill between the wells (red curve) and immediately start dropping into the left well. After a characteristic time of one full slow period, the distribution becomes essentially uniform in the two bottom wells.

\begin{figure}
 \centering
 \includegraphics[width=5.3in]{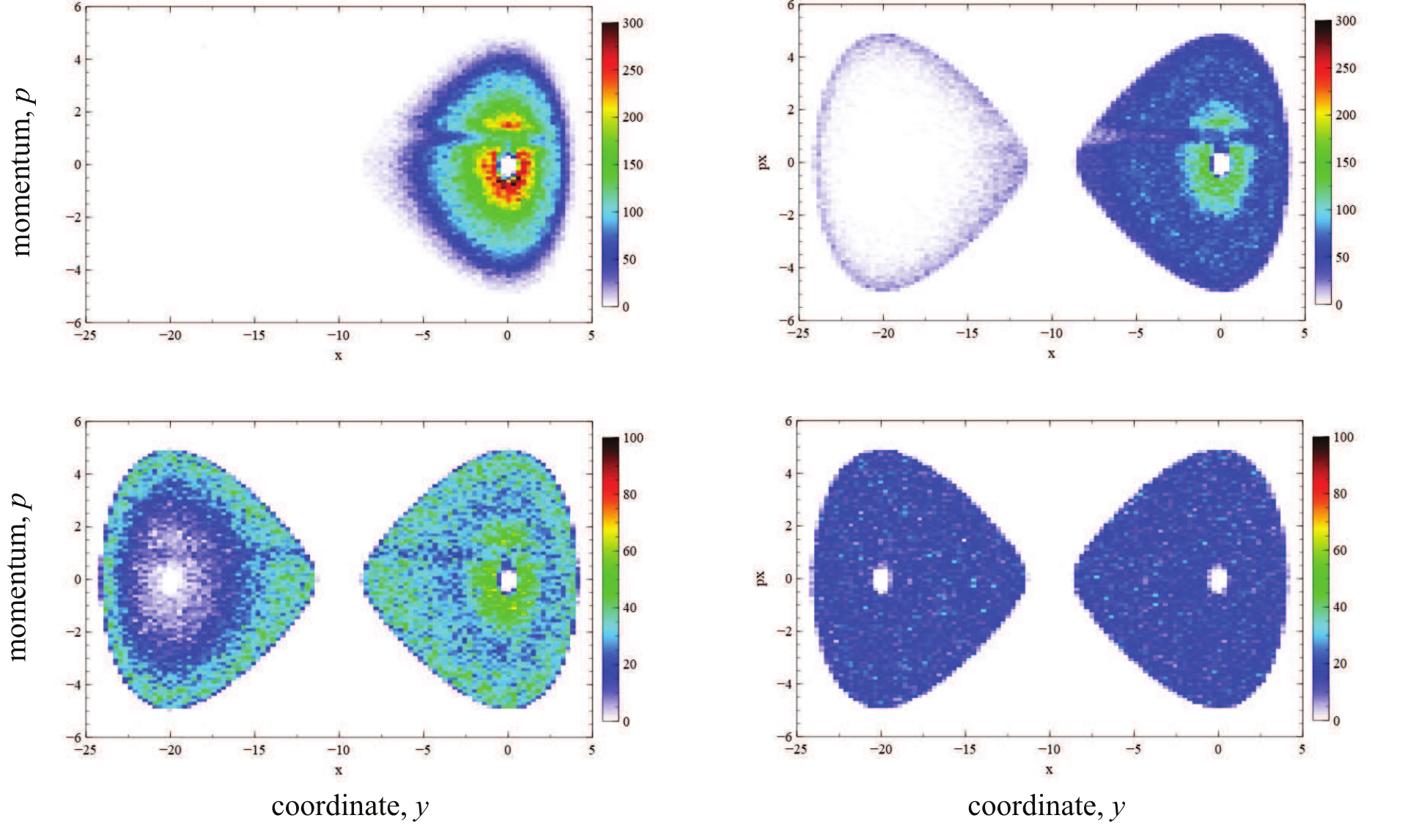}
 \caption{\label{f9} Evolution of the PDF: transport and mixing on medium time-scale. Top left: At $t\approx 100$, the particles are in the right well. Top right: at $t\approx 250$ particles fill the right well and start coming into the left well. Bottom left: at $t\approx 500$ particles fill the left well. Bottom right: at $t\approx 1000$ particles essentially uniformly fill both bottom wells. Note the differences in color scales between two top and two bottom panels.}
\end{figure}

\section{The long-time description: decay of the vortex}

The drift and diffusive spreading described above create an essentially uniform mixing over the two wells. Beyond that, over a significantly longer time time scales, the particles start drifting (diffusing) into the upper domain above the separatrix. In order to estimate the rate of transfer into the upper well, we can use the diffusion-type evolution equation for $\Psi(H,t)$. The numerical simulations indicate that the characteristic time scales of the diffusion into the upper domain occur over much longer time than the uniformization within the bottom wells.

Trapping into resonance does not play a significant role on the dynamics in the upper domain, and the resonance interactions ...

As a result, the evolution of $\Psi(H,t)$ can be described by a diffusion-type equation PDE:
\be
\frac{\pt \Psi}{\pt N} = \frac{\pt}{\pt H} \left( D(H) \frac{\pt \Psi}{\pt H} \right),
\ee
where $N$ is the number of periods on the $(x,p_x)$ plane.

The values of $D=D(H)$ is very different for $H_{min} < H < H_C$ and $H>H_C$. Additionally, there is a ``transitional'' regime $H_C < H < H_C+u^2/2 = 13$, in which trajectories intersect the resonance $4$ time on a period, which include two crossings near $x=0$.

The rate of diffusion for $H < H_C+u^2/2 = 13$ is very large compared with $H > H_C+u^2/2$. Essentially, we can set $D(H<H_C) = \infty$, and explicitly study the domain $H > H_C+u^2/2$ only (the domain $H_C < H < H_C+u^2/2$ playing a role of a transition.

For any random walk, the diffusion coefficient is equal to one half of the second moment of the corresponding distribution of the magnitude of a single step. Thus, from $\Dt H$ at a single separatrix crossing we obtain
\be
D_1(H)=\frac12 \left< \left( \Delta H(\tilde \varphi_*) \right)^2 \right>, \label{vw12}
\ee
On each period, there are two resonance crossings with the same statistical properties. Thus, we obtain a random walk in terms of $H$. Since we combine two successive resonance crossings, the total diffusion coefficient, $D(H)$, is twice as large as $D_1(H)$:
\be
D(H)=2D_1(H) = \left< \left( \Delta H(\tilde \varphi_*) \right)^2 \right>.
\label{vw13}
\ee
Converting the evolution of $\Psi$ to time, we get
\be
\frac{\pt \Psi}{\pt t} T(H) = \frac{\pt}{\pt H} \left( D(H) \frac{\pt \Psi}{\pt H} \right).
\label{vw14}
\ee
Here $T(H)$ is the period of motion, which in the upper domain is given as
\be
T(H) = 4\int\limits_{0}^{x_m} \frac{d x}{\sqrt{2H - \left( \al x^2 -1/\al \right)^2/4}}
\ee
where
\be
x_m = \frac1\al \sqrt{2\al\sqrt{2H} +1}
\ee
In the numerical simulations, for $H>12.5$, we used $D(H)$ defined by (\ref{vw13}). For $0<H<12.5$ we assumed that the particles are uniformly distributed, which corresponds to $D(H<12.5) = \infty$.

\begin{figure}
 \centering
 \includegraphics[width=2.3in]{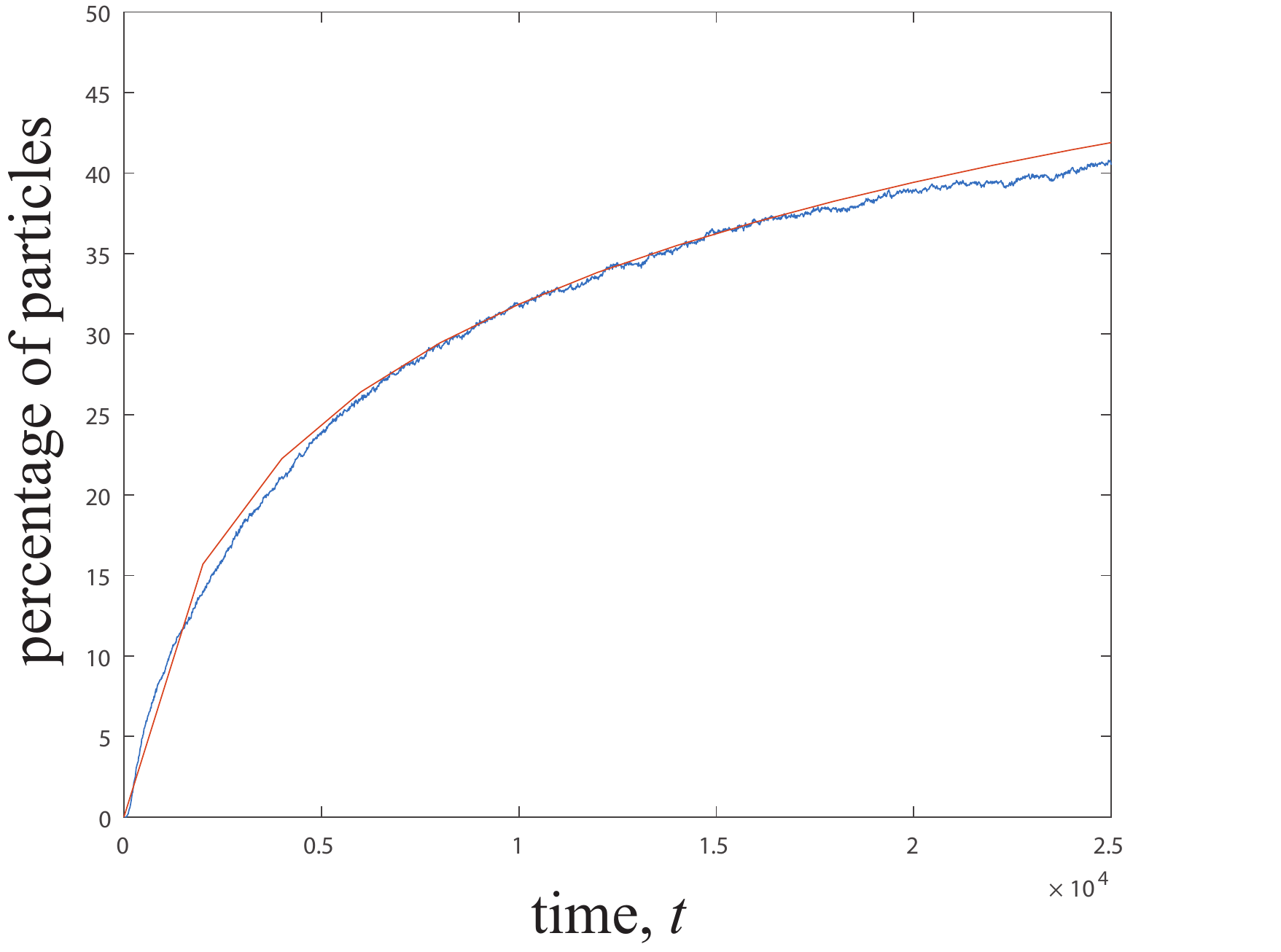}
 \caption{\label{f8} The percentage of particles in the upper domain above $H=15$. The red curve is obtained by solving PDE (\ref{vw14}) for the PDF, and the blue curve is the aggregation of explicit simulation of $20,000$ particle governed by (\ref{1}).}
\end{figure}

We compared the predictions of (\ref{vw14}) with results of explicit simulation of $20,000$ particle governed by (\ref{1}). Figure~\ref{f8} presents the amount of particles in the upper domain above $H=15$. One can see that the PDF-based description describes the leaking of particle into the upper domain.

\section{Conclusions}

In the present paper we considered a motion of charged plasma particles in a nonuniform background magnetic field in the the presence of an electrostatic wave. We proposed a setting where the nonlinear resonance phenomena create a large scale vortex is a phase space. We showed that a combination of the energy drift due to scattering at resonance and trapping (capture) at resonance creates a regular ``vorticity'' in the phase space, while the energy diffusion due to scattering at resonance create mixing. We estimated a characteristic period of the vortex and characteristic time of mixing.

\section*{Acknowledgements}
Fan Wu is fully supported by the National Natural Science Fund of China (Award No. 11702331). This material is based in part upon work supported by the National Science Foundation under Award No. CMMI-1740777 (D.V.).

\bibliographystyle{elsarticle-num}

\begin{thebibliography}{10}
\expandafter\ifx\csname url\endcsname\relax
  \def\url#1{\texttt{#1}}\fi
\expandafter\ifx\csname urlprefix\endcsname\relax\def\urlprefix{URL }\fi
\expandafter\ifx\csname href\endcsname\relax
  \def\href#1#2{#2} \def\path#1{#1}\fi

\bibitem{ONeil65}
T.~{O'Neil}, {Collisionless Damping of Nonlinear Plasma Oscillations}, Physics
  of Fluids 8 (1965) 2255--2262.
\newblock \href {http://dx.doi.org/10.1063/1.1761193}
  {\path{doi:10.1063/1.1761193}}.

\bibitem{Mazitov65}
R.~K. {Mazitov}, {Damping of plasma waves}, Journal of Applied Mechanics and
  Technical Physics 6 (1965) 22--25.
\newblock \href {http://dx.doi.org/10.1007/BF00914365}
  {\path{doi:10.1007/BF00914365}}.

\bibitem{Veltri05}
F.~{Valentini}, P.~{Veltri}, A.~{Mangeney}, {Magnetic-field effects on
  nonlinear electrostatic-wave Landau damping}, \pre 71~(1) (2005) 016402.
\newblock \href {http://dx.doi.org/10.1103/PhysRevE.71.016402}
  {\path{doi:10.1103/PhysRevE.71.016402}}.

\bibitem{Dodin&Fisch12:III}
I.~Y. {Dodin}, N.~J. {Fisch}, {Adiabatic nonlinear waves with trapped
  particles. III. Wave dynamics}, Physics of Plasmas 19~(1) (2012) 012104.
\newblock \href {http://arxiv.org/abs/1107.3075} {\path{arXiv:1107.3075}},
  \href {http://dx.doi.org/10.1063/1.3673065} {\path{doi:10.1063/1.3673065}}.

\bibitem{Benisti17:I}
D.~{B{\'e}nisti}, {Nonlocal adiabatic theory. I. The action distribution
  function}, Physics of Plasmas 24~(9) (2017) 092120.
\newblock \href {http://arxiv.org/abs/1706.03540} {\path{arXiv:1706.03540}},
  \href {http://dx.doi.org/10.1063/1.4996957} {\path{doi:10.1063/1.4996957}}.

\bibitem{Tao17:generation}
X.~Tao, F.~Zonca, L.~Chen,
  \href{http://dx.doi.org/10.1002/2017GL072624}{Identify the nonlinear
  wave-particle interaction regime in rising tone chorus generation}, \grl
  44~(8) (2017) 3441--3446.
\newblock \href {http://dx.doi.org/10.1002/2017GL072624}
  {\path{doi:10.1002/2017GL072624}}.
\newline\urlprefix\url{http://dx.doi.org/10.1002/2017GL072624}

\bibitem{Artemyev18:cnsns}
A.~V. {Artemyev}, A.~I. {Neishtadt}, V.~D. L., A.~A. {Vasiliev}, I.~Y. {Vasko},
  L.~M. {Zelenyi}, {Trapping (capture) into resonance and scattering on
  resonance: summary of results for space plasma systems}, CNSNS 65 (2018)
  111--160.
\newblock \href {http://dx.doi.org/doi:10.1016/j.cnsns.2018.05.004}
  {\path{doi:doi:10.1016/j.cnsns.2018.05.004}}.

\bibitem{Neishtadt11:mmj}
A.~{Neishtadt}, A.~{Vasiliev}, A.~{Artemyev}, {Resonance-induced surfatron
  acceleration of a relativistic particle}, Moscow Mathematical Journal 11~(3)
  (2011) 531--545.

\bibitem{Itin00}
A.~P. {Itin}, A.~I. {Neishtadt}, A.~A. {Vasiliev}, {Captures into resonance and
  scattering on resonance in dynamics of a charged relativistic particle in
  magnetic field and electrostatic wave}, Physica D: Nonlinear Phenomena 141
  (2000) 281--296.
\newblock \href {http://dx.doi.org/10.1016/S0167-2789(00)00039-7}
  {\path{doi:10.1016/S0167-2789(00)00039-7}}.

\bibitem{Neishtadt99}
A.~I. {Neishtadt}, {On Adiabatic Invariance in Two-Frequency Systems}, in
  Hamiltonian Systems with Three or More Degrees of Freedom, ed. Simï¿½ C.,
  NATO ASI Series C. Dordrecht: Kluwer Acad. Publ. 533 (1999) 193--213.
\newblock \href {http://dx.doi.org/10.1063/1.166236}
  {\path{doi:10.1063/1.166236}}.

\bibitem{Neishtadt75}
A.~{Neishtadt}, {Passage through a separatrix in a resonance problem with a
  slowly-varying parameter}, Journal of Applied Mathematics and Mechanics 39
  (1975) 594--605.
\newblock \href {http://dx.doi.org/10.1016/0021-8928(75)90060-X}
  {\path{doi:10.1016/0021-8928(75)90060-X}}.

\bibitem{bookAKN06}
V.~I. {Arnold}, V.~V. {Kozlov}, A.~I. {Neishtadt}, Mathematical Aspects of
  Classical and Celestial Mechanics, 3rd Edition, Dynamical Systems III.
  Encyclopedia of Mathematical Sciences, Springer-Verlag, New York, 2006.

\bibitem{Artemyev15:pop:probability}
A.~V. {Artemyev}, A.~A. {Vasiliev}, D.~{Mourenas}, A.~I. {Neishtadt}, O.~V.
  {Agapitov}, V.~{Krasnoselskikh}, {Probability of relativistic electron
  trapping by parallel and oblique whistler-mode waves in Earth's radiation
  belts}, Physics of Plasmas 22~(11) (2015) 112903.
\newblock \href {http://dx.doi.org/10.1063/1.4935842}
  {\path{doi:10.1063/1.4935842}}.

\bibitem{Artemyev17:pre}
A.~V. {Artemyev}, A.~I. {Neishtadt}, A.~A. {Vasiliev}, D.~{Mourenas},
  {Probabilistic approach to nonlinear wave-particle resonant interaction},
  \pre 95~(2) (2017) 023204.
\newblock \href {http://dx.doi.org/10.1103/PhysRevE.95.023204}
  {\path{doi:10.1103/PhysRevE.95.023204}}.

\end{thebibliography}

\end{document}